\documentclass[a4paper,12pt]{article}

\pdfoutput=1
\usepackage{amsmath}
\usepackage{amssymb}
\usepackage{amsfonts}
\usepackage{mathrsfs}
\usepackage{bbm}
\usepackage{graphicx,subfigure,booktabs}
\usepackage[numbers,sort&compress]{natbib}
\usepackage{verbatim}
\usepackage{color}
\usepackage{ulem}
\usepackage{setspace}
\usepackage{multirow}
\usepackage{url}

\usepackage[utf8]{inputenc}

\usepackage[colorlinks,
linkcolor=black,
filecolor=black,
anchorcolor=black,
urlcolor=black,
citecolor=blue,
bookmarks=false,
]{hyperref}

\usepackage[hyphenbreaks]{breakurl}

\usepackage{fancyhdr}

\numberwithin{equation}{section}

\newlength{\dinwidth}
\newlength{\dinmargin}
\allowdisplaybreaks

\begin{document}



\title{The $W\ell\nu$-vertex corrections to W-boson mass in the R-parity violating MSSM}

\author{
Min-Di Zheng${}^{1,2}$\footnote{zhengmd5@mail.sysu.edu.cn}\,\,,
Feng-Zhi Chen${}^{2,3}$\footnote{chenfzh25@mail.sysu.edu.cn}\,\,,
and 
Hong-Hao Zhang${}^{2}$\footnote{Correspondence: zhh98@mail.sysu.edu.cn}\\[12pt]
\small ${}^{1}$ School of Physics and Electronic Information,\\ [-0.2cm]
\small Shangrao Normal University, Shangrao 334001, China \\[-0.2cm]
\small ${}^{2}$ School of Physics, Sun Yat-Sen University, Guangzhou 510275, China\\ [-0.2cm]
\small ${}^{3}$ Key Laboratory of Quark and Lepton Physics~(MOE),\\ [-0.2cm]
\small Central China Normal University, Wuhan 430079, China
}

\date{}
\maketitle
\vspace{-0.2cm}

\begin{abstract}
{\noindent}Inspired by the astonishing $7\sigma$ discrepancy between the recent CDF-II measurement and the standard model prediction on the mass of $W$-boson, we investigate the $\lambda'$-corrections to the vertex of $\mu\to\nu_\mu e\bar{\nu_e}$ decay in the context of the $R$-parity violating minimal supersymmetric standard model. These corrections can raise the $W$-boson mass independently. Combined with recent $Z$-pole and kaon decay measurements, $m_W \lesssim 80.37$~GeV can be reached. We find that these vertex corrections cannot explain the CDF result entirely at the $2\sigma$ and even $3\sigma$ levels. However, these corrections together with the oblique contributions can be accordant with the CDF-II result and relevant bounds at the $3\sigma$ level.
\end{abstract}
\newpage

\section{Introduction}\label{sec:introduction}

In the past decades, the observation of striking agreement between the standard model (SM) predictions and the experimental results in a vast number of particle interactions has shown up the powerful predicted capacity of the SM. However, the SM is not the final answer to the particle physics, as it is unable to explain several phenomena, including the matter-antimatter asymmetry, the origin of neutrino mass, the hierarchy problem, and the candidate of dark matter. These strongly call for some new physics (NP) beyond the SM.  Although no up-to-date direct evidence shows that the NP exists, there are still indirect ways, e.g., studying the loop-effects of NP on low-energy processes or electroweak observables, like the precision measurement of the $W$-boson mass.

Recently, the Collider Detector at Fermilab (CDF) collaboration at Tevatron reported a high precision measurement on the mass of $W$-boson with the CDF-II detector. The measured value is given by $m_W^{\rm CDF}=80.4335\pm0.0094~\mathrm{GeV}$~\cite{CDF:2022hxs} with better precision than all other previous measurements and 
is $7\sigma$ from the SM prediction $m_W^\mathrm{SM}=80.357\pm0.006~\mathrm{GeV}$~\cite{Awramik:2003rn}.
If the measurement is confirmed in the future, such an astonishing tension will undoubtedly be a strong challenge to the SM.
After this exciting $m_W^\mathrm{CDF}$ reported, plentiful theoretical researches~\cite{Fan:2022dck,Zhu:2022tpr,Lu:2022bgw,Athron:2022qpo,Yuan:2022cpw,Strumia:2022qkt,Yang:2022gvz,deBlas:2022hdk,Du:2022pbp,Tang:2022pxh,Cacciapaglia:2022xih,Blennow:2022yfm,Arias-Aragon:2022ats,Zhu:2022scj,Sakurai:2022hwh,Fan:2022yly,Liu:2022jdq,Lee:2022nqz,Cheng:2022jyi,Song:2022xts,Bagnaschi:2022whn,Paul:2022dds,Bahl:2022xzi,Asadi:2022xiy,DiLuzio:2022xns,Athron:2022isz,Gu:2022htv,Heckman:2022the,Babu:2022pdn,Heo:2022dey,Du:2022brr,Cheung:2022zsb,Crivellin:2022fdf,Endo:2022kiw,Biekotter:2022abc,Balkin:2022glu} have emerged in a short time.  

Before this profound result, there are already some anomalies indicating the clues of NP, e.g., the recent average values of the observables $R_{D^{(*)}}$, reported by the Heavy Flavor Averaging Group~\cite{HFLAV:2019otj,HFLAV:2022pwe,HFLAV:2022fall}, are about $3.2\sigma$ away from the corresponding SM predictions~\cite{Bigi:2016mdz,Bernlochner:2017jka,Bigi:2017jbd,Jaiswal:2017rve,Bordone:2019vic,Gambino:2019sif,BaBar:2019vpl,Martinelli:2021onb,FermilabLattice:2021cdg}, considering the $R_{D}$ and $R_{D^\ast}$ total correlation $-0.29$. To explain these anomalies, there are numerous phenomenological studies combined with the $m_W^{\rm CDF}$ measurement in different models (e.g., see Refs.~\cite{Cheung:2022zsb,Bhaskar:2022vgk,Li:2022gwc,Chowdhury:2022dps,Allanach:2022bik,Zheng:2022ssr}). In this work, we utilize the minimal supersymmetric standard model (MSSM) extended by the $R$-parity violation (RPV), especially including the $\lambda'\hat L \hat Q \hat D$ superpotential term, which can explain the $B$-physics anomalies in the neutral current\footnote{Some anomalies are observed in the $b\to s\mu^+\mu^-$ decays include $P'_5$~\cite{Aaij:2020nrf}, the branching fraction of $ B_s \to \phi \mu^+ \mu^{-}$~\cite{LHCb:2021zwz}, etc. The ratios $R_{K^{(\ast)}}$ in the $b\to s\ell^+\ell^-$ ($\ell=e,\mu$) processes, have been reported recently by the LHCb Collaboration~\cite{LHCb:2022zom} that they are in agreement with the SM predictions, and this new result overturns the previous ones which show anomalies in $R_{K^{(\ast)}}$~\cite{LHCb:2017avl,LHCb:2019hip,LHCb:2021trn}.} or/and the charged one (see, e.g., Refs.~\cite{Altmannshofer:2017poe,Deshpand:2016cpw,Trifinopoulos:2018rna,Altmannshofer:2020axr}). Thus, further investigations on this framework for the $m_W^{\rm CDF}$ explanation are necessary. Although it is found that the MSSM can provide some parameter points which can raise $m_W$ into the $2\sigma$ accordance region~\cite{Yang:2022gvz}, mainly through bosonic self-energy contributions relevant to the oblique corrections~\cite{Peskin:1990zt,Peskin:1991sw}, the stop mass in the solution with $m_{\tilde{t}}\lesssim 1$~TeV is not suit for general collider search scenarios. Thus, it is also worth studying other corrections to $m_W$ in the extended MSSM framework, considering the general bounds for colored sparticle masses at the Large Hadron Collider (LHC). Above all, we will study corrections to the vertex $W\ell\nu$ from the $R$-parity violating interaction $\lambda'\hat L \hat Q \hat D$ and get an enhancement to $m_W$, which is independent of the oblique corrections.

This paper is organized as follows. In section~\ref{sec:vertex}, we introduce the vertex corrections to the $W$-boson mass in the MSSM framework extended by RPV. Then, we show the numerical results and discussions in section~\ref{sec:numeric}. Our conclusions are presented in section~\ref{sec:conclusion}.

\section{The contribution to $m_W$ from the R-parity violating MSSM}\label{sec:vertex}
As we know, the $W$-boson mass can be determined from the muon decay with the relation (see, e.g., Refs.~\cite{Heinemeyer:2004gx,Domingo:2011uf,Heinemeyer:2013dia})
\begin{align}\label{eq:mwmass}
\frac{m^2_W}{m^2_Z}=\frac{1}{2}+\sqrt{\frac{1}{4}-\frac{\pi \alpha}{\sqrt{2} G_\mu m^2_Z} (1+\Delta r)},
\end{align}
which comprises the three precise inputs, the $Z$-boson mass $m_Z$, the Fermi constant $G_\mu$, and the fine structure constant $\alpha$. Here the one-loop corrections to $\Delta r$ can be expressed as 
\begin{align}\label{eq:delta_r}
\Delta r=\Delta r^{\rm SM}+h^s+h^v+h^b,
\end{align}
where the SM part $\Delta r^{\rm SM}$ is derived first in Refs.~\cite{Sirlin:1980nh,Marciano:1980pb}. Within the NP part, the self-energy of the renormalized $W$-boson is denoted by $h^s$, and the vertex and box corrections to the $\mu\to\nu_\mu e\bar{\nu_e}$ decay are denoted by $h^v$ and $h^b$, respectively. In the MSSM, the pure squarks (sleptons) only engage the self-energy sector at the one-loop level. The corrections to the vertex and box involve charginos and neutralinos. Among these one-loop contributions in the MSSM, the dominant contribution to $m_W$ is the one-loop diagrams involving pure squarks. This dominant part in $h^s$ can be expressed by~\cite{Heinemeyer:2004gx}
\begin{align}
(h^s)_{\rm dom}=& -\frac{3G_\mu \cos\theta^2_W}{8\sqrt{2}\sin\theta^2_W\pi^2} \Big[
 - \sin^2\theta_{\tilde{t}}\cos^2 \theta_{\tilde{t}} F_0 (m^2_{\tilde{t}_1}, m^2_{\tilde{t}_2})
 - \sin^2\theta_{\tilde{b}}\cos^2 \theta_{\tilde{b}} F_0 (m^2_{\tilde{b}_1}, m^2_{\tilde{b}_2}) \notag\\
& + \cos^2 \theta_{\tilde{t}}\cos^2 \theta_{\tilde{b}} F_0 (m^2_{\tilde{t}_1}, m^2_{\tilde{b}_1})
  + \cos^2 \theta_{\tilde{t}}\sin^2 \theta_{\tilde{b}} F_0 (m^2_{\tilde{t}_1}, m^2_{\tilde{b}_2}) \notag\\ 
& + \sin^2 \theta_{\tilde{t}}\cos^2 \theta_{\tilde{b}} F_0 (m^2_{\tilde{t}_2}, m^2_{\tilde{b}_1}) 
  + \sin^2 \theta_{\tilde{t}}\sin^2 \theta_{\tilde{b}} F_0 (m^2_{\tilde{t}_2}, m^2_{\tilde{b}_2})
\Big],
\end{align} 
where $\theta_W$ is the Weinberg angle and the definition of mixing angle $\theta_{\tilde{q}}$ is referred to Ref.~\cite{Heinemeyer:2004gx} and the function $F_0(x,y)=x+y-\frac{2xy}{x-y} \log\frac{x}{y}$ with the extra properties $F_0(m^2,m^2)=0$ and $F_0(m^2,0)=m^2$. 
Thus, one can see that $h^s$ is sensitive to the mass splitting between the isospin partners due to the factor $\cos^2 \theta_{\tilde{t}} \cos^2 \theta_{\tilde{b}}$.
Obviously, $h^s$ can be negligible when the soft breaking masses $M_{\tilde{Q}_i}$ are sufficiently heavy compared to the chiral mixing. In this work, we focus on the vertex corrections $h^v$ affected by the $\lambda'$-coupling in the $R$-parity violating MSSM (RPV-MSSM) and can omit $h^s$ and $h^b$ in the particular scenario.

In RPV-MSSM, the $\lambda'$-superpotential term ${\cal W}=\lambda'_{ijk} \hat L_i \hat Q_j \hat D_k$ leads to the related Lagrangian in the mass basis
\begin{align}
{\cal L}^{\rm LQD} = &\lambda'_{ijk} \left(\tilde{\nu}_{Li} \bar{d}_{Rk} d_{Lj} + \tilde{d}_{Lj} \bar{d}_{Rk} \nu_{Li} + \tilde{d}_{Rk}^\ast \bar{\nu}_{Li}^c d_{Lj} \right)   \notag \\ 
&- \tilde{\lambda}'_{ijk} \left( \tilde{l}_{Li} \bar{d}_{Rk} u_{Lj} + \tilde{u}_{Lj} \bar{d}_{Rk} l_{Li} + \tilde{d}_{Rk}^\ast \bar{l}_{Li}^c u_{Lj} \right) + {\rm h.c.}, 
\end{align}
where the generation indices $i,j,k=1,2,3$, while the color ones are  omitted, and ``$c$'' indicates the charge conjugated fermions. In this paper, all the repeated indices are defaulted to be summed over unless otherwise stated. The relation between $\lambda'$ and $\tilde{\lambda}'$ is $\tilde{\lambda}'_{ijk}=\lambda'_{ij'k} K^{\ast}_{jj'}$ with $K$ being the Cabibbo–Kobayashi–Maskawa (CKM) matrix. In this work, we restrict the index $k$ of the superfield $\hat D_k$ to the single value $3$.

\begin{figure}[htbp]
	\centering
	\includegraphics[width=0.5\textwidth]{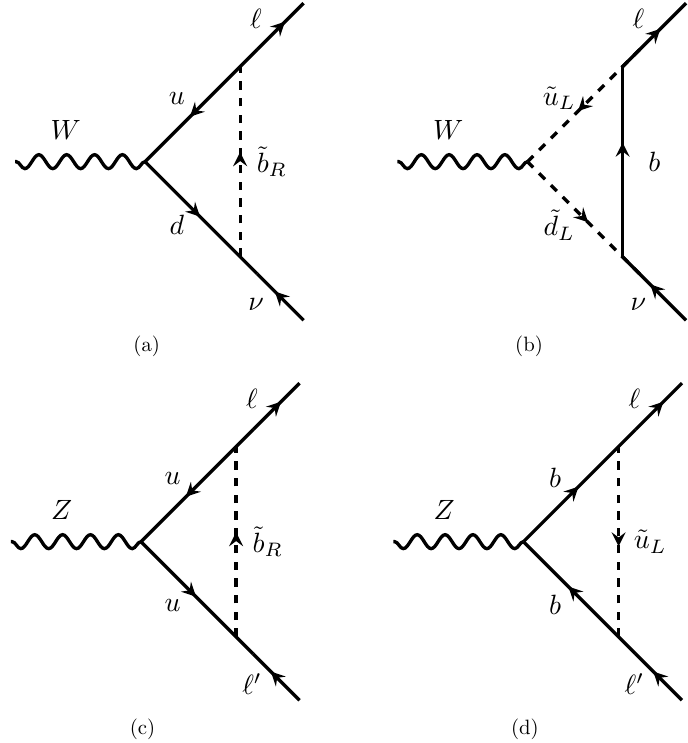}
	\caption{The diagrams of the $W\ell\nu$-vertex and $Z\ell\ell^{(')}$-vertex (shown as examples) in the RPV-MSSM.}
	\label{fig:diagrams}
\end{figure}  

Including the one-loop contribution from the RPV-MSSM, the $W \ell_l \nu_{i}$-vertex is described by the following Lagrangian
\begin{align}\label{eq:LagWlnu}
{\cal L}^W_{\rm eff}=\frac{g}{\sqrt{2}} \bar{\ell_l} \gamma^\mu P_L (\delta_{li}+h_{li}) \nu_i W^{-}_\mu + {\rm h.c.},
\end{align}
where $g$ is the ${\rm SU}(2)_L$ gauge coupling, and the correction part $h_{li}$ from the $\lambda'$-contributions is given by (as the analogy to the formula in Ref.~\cite{Arnan:2019olv})
\begin{align}\label{eq:hprime}
h'_{li} = -\frac{3}{64\pi^2} x_t f_W(x_t) \tilde{\lambda}'^{\ast}_{l33} \tilde{\lambda}'_{i33},
\end{align}
where $x_t \equiv m^2_t/m^2_{\tilde{b}_R}$ and the loop function $f_W(x) \equiv \frac{1}{x-1} + \frac{(x-2) \log x}{(x-1)^2}$ and other non-dominant parts are eliminated. This dominant contribution is from the vertex engaged by the right-handed sbottom $\tilde{b}_R$ (figure~\ref{fig:diagrams}a) while the vertex involving left-handed squarks (figure~\ref{fig:diagrams}b) provides non-dominant effects and can be eliminated. 
Then, we consider the $\lambda'$-correction only to the $W\mu\nu$-vertex or to the $We\nu$-vertex at a time. This can be easily achieved by setting one of the couplings $(\tilde{\lambda}'_{133}, \tilde{\lambda}'_{233})$ dominant while neglecting the rest. 
Given this ``single coefficient dominance'' scenario, the $\lambda'$-corrections to the $\mu\to\nu_\mu e\bar{\nu_e}$ box also vanish,\footnote{In this scenario, the $\lambda'$-contributions to the $\mu\to\nu_i e\bar{\nu_i}$ through $Z$ penguin vanish as well.} then the one-loop $\lambda'$-contribution to $\Delta r$ only comes from $h'_{aa}$ (the index $a$ here is restricted to $1$ or $2$ at a time).

Given the purpose of this work is to investigate that to what degree, the pure $\lambda'$ contribution, $h'_{aa}$, can accommodate the new $W$-boson mass data. We can further write down the prediction of the $W$-boson from the pure-$\lambda'$ contributions\footnote{There are always contributions from the original MSSM framework, while we can set  sufficiently heavy masses of left-handed squarks, sleptons, and gauginos in soft breaking terms to screen these effects.} as
\begin{align}\label{eq:mwNew}
(m^{\lambda'}_W/{\rm GeV})=80.357-15.6387 h'_{aa}=80.357+0.0743 x_t f_W(x_t) 
\left| \tilde{\lambda}'_{a33} \right|^2
\end{align}
with Eqs.~\eqref{eq:mwmass} and \eqref{eq:hprime}.
It is clear from Eq.~\eqref{eq:mwNew} that the right-handed sbottom mass $m_{\tilde{b}_R}$ and the coupling $\tilde{\lambda}'_{a33}$ are related to $\lambda'$-correction of $m_W$.

\section{Numerical results and discussions}\label{sec:numeric}
In this section, we investigate the explanation of $m_W^{\rm CDF}$ combined with the relevant constraints. At first, we concentrate on the pure $\lambda'$-effects assuming the soft breaking masses of gauginos and left-handed squarks (sleptons) are sufficiently heavy, and then, only the model parameters $(\tilde{\lambda}'_{a33},m_{\tilde{b}_R})$ are involved. 
If the pure $\lambda'$-contribution (see Eq.~\eqref{eq:mwNew}) can explain the new $W$-boson mass at the $2\sigma$ level, we need $h'_{aa}$ to fulfill $-6.34<h'_{aa}\times 10^3<-3.47$. With Eq.~\eqref{eq:hprime}, this bound provides 
\begin{align}\label{eq:mwbound2sigma}
0.7313< x_t f_W(x_t) \left| \tilde{\lambda}'_{a33} \right|^2
<1.3357.
\end{align}
Also, $-6.34<h'_{aa}\times 10^3<-3.47$ let the ratio 
\begin{align} 
R^W_{\rm NP/SM}\equiv\Gamma(W\to\ell_a\nu_a)_{\rm NP}/\Gamma(W\to\ell_a\nu_a)_{\rm SM}
\end{align}
stay in the region
\begin{align}\label{eq:Wratio}
-12.7<(R^W_{\rm NP/SM}-1)\times 10^3<-6.9,
\end{align}
because $R^W_{\rm NP/SM}$ can be calculated as $1+2h'_{aa}$, with Eq.~\eqref{eq:LagWlnu}.
Then, we compare Eq.~\eqref{eq:Wratio} with the $W$-boson partial width ratios $R^W_{l/l'}\equiv\Gamma(W\to l\nu)/\Gamma(W\to l'\nu)$, and their experimental results are given as $R^W_{\mu/e}=0.996\pm 0.008$, $R^W_{\tau/\mu}=1.008\pm 0.031$, and $R^W_{\tau/e}=1.043\pm 0.024$~\cite{ParticleDataGroup:2020ssz}. It is found that the $m^{\rm}_W$ explanation demands much stronger bounds, whenever the NP exists in the $\mu$ or $e$ channel (the $\tau$ flavor is assumed decoupled with the NP for simplicity).

As shown in figure~\ref{fig:diagrams}c and d, the NP effects on the $W\ell\nu$-vertex will also inevitably affect the $Z$-vertex. The $Z$-boson partial width ratios $R^Z_{l/l'}\equiv\Gamma(Z\to ll)/\Gamma(Z\to l'l')$ are measured as $R^Z_{\mu/e}=1.0001\pm 0.0024$, $R^Z_{\tau/\mu}=1.0010\pm 0.0026$, and $R^Z_{\tau/e}=1.0020\pm 0.0032$~\cite{ParticleDataGroup:2020ssz}, which all constrain the coupling $g_{\ell_L}$ in the effective Lagrangian
\begin{align}
{\cal L}^Z_{\rm eff}=\frac{g}{\cos\theta_W} \bar{\ell}_i \gamma^\mu \left[g_{\ell_L}^{ij} P_L + g_{\ell_R}^{ij} P_R \right] \ell_j Z_\mu,
\end{align}
where $g_{\ell_L}^{ij}=\delta^{ij}g_{\ell_L}^{\rm SM}+\delta g_{\ell_L}^{ij}+\delta g_{\ell_L}^{\prime ij}$ and $g_{\ell_R}^{ij}=\delta^{ij}g_{\ell_R}^{\rm SM}$, with $g_{\ell_L}^{\rm SM}=-\frac{1}{2}+\sin^2\theta_W$ and $g_{\ell_R}^{\rm SM}=\sin^2\theta_W$. 
The formulas of $\delta g_{\ell_L}^{ij}$ and $\delta g_{\ell_L}^{\prime ij}$ are~\cite{Earl:2018snx}
\begin{align}
(32\pi^2) \delta g_{\ell_L}^{ij} =& 3 \tilde{\lambda}'_{j33} \tilde{\lambda}'^{\ast}_{i33} \biggl\{-x_t (1 + \log x_t) 
+ \frac{m_Z^2}{18 m^2_{\tilde{b}_R}}\biggl[(11 - 10 \sin^2\theta_W) \notag \\ 
&+ (6 - 8 \sin^2\theta_W)\log x_t + \frac{1}{10}(-9 + 16 \sin^2\theta_W)\frac{m_Z^2}{m_t^2} \biggr] \biggl\},   \notag\\
(32\pi^2) \delta g_{\ell_L}^{\prime ij}=& \tilde{\lambda}'_{j33} \tilde{\lambda}'^{\ast}_{i33} \frac{m^2_Z}{m^2_{\tilde{t}_L}} \biggl[
\left(-\frac{2}{3} \sin^2\theta_W \right) \left( \log\left( \frac{m^2_Z}{m^2_{\tilde{t}_L}} \right) - i \pi - \frac{1}{2}\right) + \left( -\frac{1}{6} + \frac{1}{9} \sin^2\theta_W \right) \biggr].
\end{align}
Here we can define $B^{ij} \equiv (32\pi^2) (\delta g_{\ell_L}^{ij}+\delta g_{\ell_L}^{\prime ij})$ and further get the bound $|B^{aa}|<0.35(0.53)$ at the $2(3)\sigma$ level. Given that the mass of $\tilde{t}_L$ is set sufficiently heavy, the $\delta g_{\ell_L}^{\prime ij}$ part can be eliminated.

As to the invisible $Z$-decay, this model can also make loop-level  contributions to the $Z\to \nu\bar{\nu}$, i.e., $\ell$ exchanged with $\nu$ and $u$($\tilde{u}_L$) exchanged by $d$($\tilde{d}_L$) in figure~\ref{fig:diagrams}c, d. Then, the effective number of light neutrinos $N_\nu$, which is defined by $\Gamma_{\rm inv}=N_\nu \Gamma_{\nu\bar{\nu}}^{\rm SM}$~\cite{ALEPH:2005ab}, will constrain the couplings via
\begin{align}
N_\nu=\left| 1+\frac{\delta g_{\nu}^{aa}+\delta g_{\nu}^{\prime aa}}{\delta g_{\nu}^{\rm SM}} \right|^2+2,
\end{align}   
where the coupling $\delta g_{\nu}^{\rm SM}=\frac{1}{2}$ and the formulas of $\delta g_{\nu}^{(\prime)ij}$ is given by
\begin{align}
(32\pi^2) \delta g_{\nu}^{ij} =& \lambda'_{j33} \lambda'^{\ast}_{i33}
\frac{m^2_Z}{m^2_{\tilde{b}_R}}
\biggl[ \left(-1+\frac{2}{3} \sin^2\theta_W \right) \left( \log \left( \frac{m^2_Z}{m^2_{\tilde{b}_R}} \right) - i \pi - \frac{1}{3}  \right) + \left( -\frac{1}{12} + \frac{4}{9} \sin^2\theta_W \right) \biggr],
\notag \\ 
(32\pi^2) \delta g_{\nu}^{\prime ij}=& \lambda'_{j33} \lambda'^{\ast}_{i33} \frac{m^2_Z}{m^2_{\tilde{b}_L}} \biggl[
\left( \frac{1}{3} \sin^2\theta_W \right) \left( \log\left( \frac{m^2_Z}{m^2_{\tilde{b}_L}} \right) - i \pi - \frac{1}{2}\right) + \left( -\frac{1}{6} + \frac{4}{9} \sin^2\theta_W \right) \biggr].
\end{align}
Then the measurement $N_\nu^{\rm exp}=2.9840(82)$~\cite{ALEPH:2005ab} will make constraints.

Except the purely leptonic decays of $W/Z$ boson, the $\mu\to e\bar{\nu}_e\nu_\mu$ and $\tau\to\ell\bar{\nu}_{\ell}\nu_\tau$ decays, which contain the $W\ell\nu$-vertex, should also be considered. The fraction ratios ${\cal B}(\tau\to\mu\bar{\nu}_\mu\nu_\tau)/{\cal B}(\tau\to e\bar{\nu}_e\nu_\tau)$, ${\cal B}(\tau\to e\bar{\nu}_e\nu_\tau)/{\cal B}(\mu\to e\bar{\nu}_e\nu_\mu)$, and ${\cal B}(\tau\to\mu\bar{\nu}_\mu\nu_\tau)/{\cal B}(\mu\to e\bar{\nu}_e\nu_\mu)$ 
make the bounds~\cite{Bryman:2021teu} as
\begin{align}\label{eq:cons_lepton}
\frac{1+\delta^{\rm a 2} h'_{\mu\mu}}
{1+\delta^{\rm a 1} h'_{ee}}&=1.0018(14),  \notag\\
\frac{1}
{1+\delta^{\rm a 2} h'_{\mu\mu}}&=1.0010(14),  \notag\\
\frac{1}
{1+\delta^{\rm a 1} h'_{ee}}&=1.0029(14).
\end{align}
Due to that $h'_{aa}\leqslant 0$, Eq.~\eqref{eq:cons_lepton} induces the $2(3)\sigma$ bounds $-4.6(-6.0)<h'_{ee}\times 10^3 \leqslant 0$ or $-1(-2.4)<h'_{\mu\mu}\times 10^3 \leqslant 0$ for $a$ restricted to $1$ or $2$, respectively. Similarly, the decays of kaons and pions also make the bounds, and the most stringent ones~\cite{Bryman:2021teu} induce $-1.4(-3.2)<h'_{ee}\times 10^3 \leqslant 0$ or $-0.8(-1.7)<h'_{\mu\mu}\times 10^3 \leqslant 0$ at the $2(3)\sigma$ level, which are provided by the constraints in the $K^+\to\ell^+\nu(\gamma)$ and $\pi\to\ell\nu(\gamma)$ decays, respectively. However, the $2\sigma$-level $m^{\rm CDF}_W$ explanation demands  $-6.34<h'_{aa}\times 10^3<-3.47$ as mentioned before, and thus, one can see exactly the $2\sigma$-level explanation is already excluded by these decays of kaons and pions, while we can still investigate the degree of the NP raising $m_W$. Given that $h'_{\mu\mu}$ is constrained more strongly than $h'_{ee}$, in the following we focus the NP in the $e$ flavor. Thus, we set $\lambda'_{133}$ dominant and other $\lambda'$ couplings negligible. Then with the CKM rotation, $\tilde{\lambda}'_{ijk}=\lambda'_{ij'k} K^{\ast}_{jj'}$, the nonzero $\tilde{\lambda}'$ couplings $|\tilde{\lambda}'_{133}| \approx |\lambda'_{133}|$, $|\tilde{\lambda}'_{123}| \approx 0.04 |\lambda'_{133}|$, and 
$|\tilde{\lambda}'_{113}| \approx 10^{-3} |\lambda'_{133}|$. As we consider the real number $\lambda'_{133}$ varying in $0\leqslant \lambda'_{133} \leqslant 3$, it is checked that the effects on the $m^{\rm CDF}_W$ explanation and constraints from the couplings $\tilde{\lambda}'_{123}$ and $\tilde{\lambda}'_{113}$ are negligible.  
It is worth to mention that all the model parameters are set at the scale of around TeV. Given that only (axial-)vector currents are involved in relevant processes discussed before, the couplings of these currents are kept nearly the same when the scale runs down to the electroweak scale. 

\begin{figure}[htbp]
	\centering
	\includegraphics[width=0.95\textwidth]{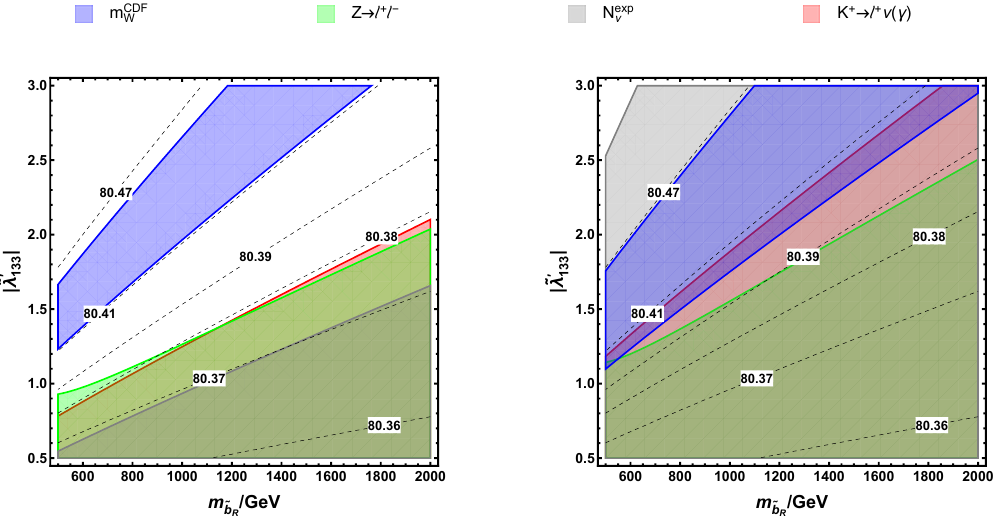}
	\caption{The regions of $(m_{\tilde{b}_R},|\tilde{\lambda}'_{133}|)$ at the $2\sigma$ (left) and $3\sigma$ (right) levels.
The pure $\lambda'$-contributions to the $m^{\rm CDF}_W$ explanation  are denoted by the blue. The areas allowed by $R^Z_{\ell/\ell'}$ and $N_\nu^{\rm exp}$ are shown by the green and gray, respectively. The areas filled by the red are allowed by the data of $K^+\to\ell^+\nu(\gamma)$ decay. 
The dashed lines express the $m^{\lambda'}_W$ value (GeV) enhanced by vertex corrections.}
	\label{fig:bound1}
\end{figure}

Combining the bounds introduced above with the $W$ mass explanation, the allowed regions are shown in figure~\ref{fig:bound1}.
The two areas allowed by $Z\to\ell\ell$ and kaon decays overlap almost entirely at the $2\sigma$ level, while the $Z\to\ell\ell$ bound is stronger at the $3\sigma$ level. The bounds of $N_\nu^{\rm exp}$ is more stringent than the former two at the $2\sigma$ level, but the loosest at the $3\sigma$ level. 
In the common region of these three observables at the $2\sigma$ level, $m^{\lambda'}_W$ can be raised to around $80.37$~GeV at most, while it cannot reach the value to explain $m^{\rm CDF}_W$ as predicted. Even at the $3\sigma$ level, there are still none common areas for $m^{\rm CDF}_W$ and bounds besides the one when $m_{\tilde{b}_R} \lesssim 600$~GeV, but this mass scale is already excluded by LHC searches~\cite{ATLAS:2021jyv,ATLAS:2015oan,CMS:2016zgb}. Therefore, we find that the pure $\lambda'$ contributions cannot fully solve the $m_W$ problem unless with other effects, e.g., the oblique corrections~\cite{Bagnaschi:2022qhb,Yang:2022gvz}. Thus, we will further study the combination explanation with the $\lambda'$-contributions and the oblique ones of the MSSM framework. 

\begin{table}[htbp]
\centering
\setlength\tabcolsep{8pt}
\renewcommand{\arraystretch}{1.3}
\begin{tabular}{|cc||cc|}
\hline
Parameters & Sets & Parameters & Sets \\
\hline
$\tan\beta$ & $15$ & $M_{\tilde{L}_{1,2,3}}=M_{\tilde{E}_{1,2,3}}$ & $2000$\\
$\mu$       & $1000$ & $M_{\tilde{Q}_{1,2}}=M_{\tilde{U}_{1,2}}=M_{\tilde{D}_{1,2}}$ & $10^4$\\
$M_1$ & $500$ & $M_{\tilde{Q}_{3}}$ & $1500\sim3000$\\
$M_2$ & $1000$ & $M_{\tilde{U}_3}$ & $1500\sim10^4$\\
$M_3$ & $5000$ & $M_{\tilde{D}_3}$ & $1300\sim3000$\\
$M_A$ & $2000$ & $A_{u,c}=A_{d,s}=A_l$ & $1500$\\
$m_t$ & $173.3$ & $A_t$, $A_b$ & $-5000\sim5000$
\\
\hline
	\end{tabular}
	\caption{The sets of parameters in the MSSM part. Parameters with mass dimension are given in GeV. The lower limits of squark masses refer to Ref.~\cite{ATLAS:2021jyv}.}
	\label{tab:parameter}
\end{table}

\begin{figure}[htbp]
	\centering
	\includegraphics[width=0.95\textwidth]{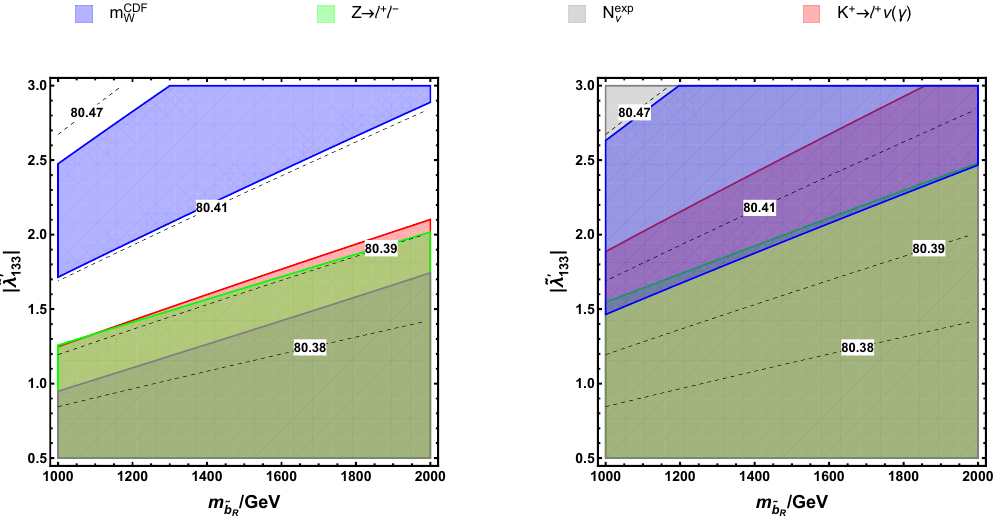}
	\caption{Same as figure~\ref{fig:bound1} but the dashed lines express the $m^{\rm NP}_W$ value (GeV) enhanced by both vertex corrections and oblique ones of the MSSM.}
	\label{fig:bound2}
\end{figure}

Different from the pure-RPV case that only parameters $(\tilde{\lambda}'_{133},m_{\tilde{b}_R})$ are focused on, in the following we further consider non-decoupled masses of stops and gauginos, and the parameters are collected in table~\ref{tab:parameter}.
Then, we utilize {\tt FeynHiggs-2.18.1}~\cite{Heinemeyer:1998yj,Heinemeyer:1998np,Degrassi:2002fi,Frank:2006yh,Hahn:2013ria,Bahl:2016brp,Bahl:2017aev,Bahl:2018qog} to calculate the loop correction of MSSM part, i.e., $h^s$, which is given as the nearly fixed  value $h^s \approx -8\times 10^{-4}$ for the parameters $M_{\tilde{Q}_{3}}$, $M_{\tilde{U}_3}$, $M_{\tilde{D}_3}$, $A_t$, and $A_b$ varying in the ranges shown in table~\ref{tab:parameter}, also keeping the mass of Higgs-like boson in $122<m_H<128$~GeV. Then, we can further set $M_{\tilde{Q}_3}=2.1$~TeV, $M_{\tilde{U}_3}=10$~TeV, and  $A_t=A_b=1.5$~TeV as the benchmark point, and write down the prediction of $m_W$ from  combined contributions as
\begin{align}
(m^{\rm NP}_W/{\rm GeV})=80.370-15.622 h'_{aa}=80.370 + 0.0742 x_t f_W(x_t) 
\left| \tilde{\lambda}'_{a33} \right|^2.
\end{align}
Then, the allowed regions are shown in figure~\ref{fig:bound2}.
One can see that $m_W$ can be raised to around $80.38$~GeV in the $2\sigma$-level allowed region of the $Z$ and kaon decays, and explaining the $W$-mass anomaly at $2\sigma$ is still unachievable. However, the $3\sigma$-level explanation is allowed by all the bounds, within the narrow overlap near the edge of $m^{\rm CDF}_W$ region.

\section{Conclusions}\label{sec:conclusion}

In this paper, inspired by the astonishing $7\sigma$ discrepancy between the CDF-II measurement and the SM prediction on the mass of $W$-boson, we performed a phenomenological analysis on the muon decay that is relevant to the $W$ mass under the framework of RPV-MSSM, to access whether such a deviation can be accommodated by this NP model. We focused on the one-loop corrections to the vertex of $\mu\to\nu_\mu e\bar{\nu_e}$ decay, assuming that the vertex correction is only affected by a single $\lambda'$ coupling in the RPV-MSSM. The numerical results shown in figure~\ref{fig:bound1}  imply that pure $\lambda'$-contributions in the RPV-MSSM are hard to accommodate the CDF measurement entirely. However, the $\lambda'$-corrections can help raise the prediction of $W$ mass to be accordant with $m^{\rm CDF}_W$ at the $3\sigma$ level when combined with the oblique corrections, which is shown in figure~\ref{fig:bound2}.

\subsection*{Acknowledgements}
We thank Chengfeng Cai and Seishi Enomoto for valuable discussions.
This work is supported in part by the National Natural Science Foundation of China under Grant Nos. 11875327 and 12275367, the Fundamental Research Funds for the Central Universities, and the Sun Yat-Sen University Science Foundation. F.C. is also supported by the CCNU-QLPL Innovation Fund (QLPL2021P01).

\subsection*{Authors’ contributions}
M.D. contributed to the study conception and design. All authors commented on previous versions of the manuscript. All authors read and approved the final manuscript.

\subsection*{Funding}
This research was supported in part by the National Natural Science Foundation of China under Grant Nos. 11875327 and 12275367, the Fundamental Research Funds for the Central Universities, and the Sun Yat-Sen University Science Foundation. F.C. is also supported by the CCNU-QLPL Innovation Fund (QLPL2021P01).

\subsection*{Availability of data and materials}
All data generated or analyzed during this study are included in this published article.

\section*{Declarations}

\subsection*{Competing interests}
The authors declare that they have no competing interests.

\subsection*{Ethics approval and consent to participate}
The authors declare they have upheld the integrity of the scientific record.

\subsection*{Consent for publication}
The authors give their consent for publication of this article.

\bibliographystyle{JHEP}
\bibliography{ref}

\providecommand{\href}[2]{#2}\begingroup\raggedright\begin{thebibliography}{10}

\bibitem{CDF:2022hxs}
{\bf CDF} Collaboration, T.~Aaltonen et~al., {\it {High-precision measurement
  of the $W$ boson mass with the CDF II detector}},  {\it Science} {\bf 376}
  (2022), no.~6589 170--176.

\bibitem{Awramik:2003rn}
M.~Awramik, M.~Czakon, A.~Freitas, and G.~Weiglein, {\it {Precise prediction
  for the W boson mass in the standard model}},  {\it Phys. Rev. D} {\bf 69}
  (2004) 053006, [\href{http://arxiv.org/abs/hep-ph/0311148}{{\tt
  hep-ph/0311148}}].

\bibitem{Fan:2022dck}
Y.-Z. Fan, T.-P. Tang, Y.-L.~S. Tsai, and L.~Wu, {\it {Inert Higgs Dark Matter
  for CDF II W-Boson Mass and Detection Prospects}},  {\it Phys. Rev. Lett.}
  {\bf 129} (2022), no.~9 091802, [\href{http://arxiv.org/abs/2204.03693}{{\tt
  arXiv:2204.03693}}].

\bibitem{Zhu:2022tpr}
C.-R. Zhu, M.-Y. Cui, Z.-Q. Xia, Z.-H. Yu, X.~Huang, Q.~Yuan, and Y.-Z. Fan,
  {\it {Explaining the GeV Antiproton Excess, GeV \ensuremath{\gamma}-Ray
  Excess, and W-Boson Mass Anomaly in an Inert Two Higgs Doublet Model}},  {\it
  Phys. Rev. Lett.} {\bf 129} (2022), no.~23 231101,
  [\href{http://arxiv.org/abs/2204.03767}{{\tt arXiv:2204.03767}}].

\bibitem{Lu:2022bgw}
C.-T. Lu, L.~Wu, Y.~Wu, and B.~Zhu, {\it {Electroweak precision fit and new
  physics in light of the W boson mass}},  {\it Phys. Rev. D} {\bf 106} (2022),
  no.~3 035034, [\href{http://arxiv.org/abs/2204.03796}{{\tt
  arXiv:2204.03796}}].

\bibitem{Athron:2022qpo}
P.~Athron, A.~Fowlie, C.-T. Lu, L.~Wu, Y.~Wu, and B.~Zhu, {\it {Hadronic
  uncertainties versus new physics for the W boson mass and Muon
  g\,\ensuremath{-}\,2 anomalies}},  {\it Nature Commun.} {\bf 14} (2023),
  no.~1 659, [\href{http://arxiv.org/abs/2204.03996}{{\tt arXiv:2204.03996}}].

\bibitem{Yuan:2022cpw}
G.-W. Yuan, L.~Zu, L.~Feng, Y.-F. Cai, and Y.-Z. Fan, {\it {Is the W-boson mass
  enhanced by the axion-like particle, dark photon, or chameleon dark
  energy?}},  {\it Sci. China Phys. Mech. Astron.} {\bf 65} (2022), no.~12
  129512, [\href{http://arxiv.org/abs/2204.04183}{{\tt arXiv:2204.04183}}].

\bibitem{Strumia:2022qkt}
A.~Strumia, {\it {Interpreting electroweak precision data including the W-mass
  CDF anomaly}},  {\it JHEP} {\bf 08} (2022) 248,
  [\href{http://arxiv.org/abs/2204.04191}{{\tt arXiv:2204.04191}}].

\bibitem{Yang:2022gvz}
J.~M. Yang and Y.~Zhang, {\it {Low energy SUSY confronted with new measurements
  of W-boson mass and muon g-2}},  {\it Sci. Bull.} {\bf 67} (2022), no.~14
  1430--1436, [\href{http://arxiv.org/abs/2204.04202}{{\tt arXiv:2204.04202}}].

\bibitem{deBlas:2022hdk}
J.~de~Blas, M.~Pierini, L.~Reina, and L.~Silvestrini, {\it {Impact of the
  Recent Measurements of the Top-Quark and W-Boson Masses on Electroweak
  Precision Fits}},  {\it Phys. Rev. Lett.} {\bf 129} (2022), no.~27 271801,
  [\href{http://arxiv.org/abs/2204.04204}{{\tt arXiv:2204.04204}}].

\bibitem{Du:2022pbp}
X.~K. Du, Z.~Li, F.~Wang, and Y.~K. Zhang, {\it {Explaining The Muon $g-2$
  Anomaly and New CDF II W-Boson Mass in the Framework of (Extra)Ordinary Gauge
  Mediation}},  \href{http://arxiv.org/abs/2204.04286}{{\tt arXiv:2204.04286}}.

\bibitem{Tang:2022pxh}
T.-P. Tang, M.~Abdughani, L.~Feng, Y.-L.~S. Tsai, and Y.-Z. Fan, {\it {NMSSM
  neutralino dark matter for $W$-boson mass and muon $g-2$ and the promising
  prospect of direct detection}},  \href{http://arxiv.org/abs/2204.04356}{{\tt
  arXiv:2204.04356}}.

\bibitem{Cacciapaglia:2022xih}
G.~Cacciapaglia and F.~Sannino, {\it {The W boson mass weighs in on the
  non-standard Higgs}},  {\it Phys. Lett. B} {\bf 832} (2022) 137232,
  [\href{http://arxiv.org/abs/2204.04514}{{\tt arXiv:2204.04514}}].

\bibitem{Blennow:2022yfm}
M.~Blennow, P.~Coloma, E.~Fern\'andez-Mart\'\i{}nez, and M.~Gonz\'alez-L\'opez,
  {\it {Right-handed neutrinos and the CDF II anomaly}},  {\it Phys. Rev. D}
  {\bf 106} (2022), no.~7 073005, [\href{http://arxiv.org/abs/2204.04559}{{\tt
  arXiv:2204.04559}}].

\bibitem{Arias-Aragon:2022ats}
F.~Arias-Arag\'on, E.~Fern\'andez-Mart\'\i{}nez, M.~Gonz\'alez-L\'opez, and
  L.~Merlo, {\it {Dynamical Minimal Flavour Violating inverse seesaw}},  {\it
  JHEP} {\bf 09} (2022) 210, [\href{http://arxiv.org/abs/2204.04672}{{\tt
  arXiv:2204.04672}}].

\bibitem{Zhu:2022scj}
B.-Y. Zhu, S.~Li, J.-G. Cheng, R.-L. Li, and Y.-F. Liang, {\it {Using gamma-ray
  observation of dwarf spheroidal galaxy to test a dark matter model that can
  interpret the W-boson mass anomaly}},
  \href{http://arxiv.org/abs/2204.04688}{{\tt arXiv:2204.04688}}.

\bibitem{Sakurai:2022hwh}
K.~Sakurai, F.~Takahashi, and W.~Yin, {\it {Singlet extensions and W boson mass
  in light of the CDF II result}},  {\it Phys. Lett. B} {\bf 833} (2022)
  137324, [\href{http://arxiv.org/abs/2204.04770}{{\tt arXiv:2204.04770}}].

\bibitem{Fan:2022yly}
J.~Fan, L.~Li, T.~Liu, and K.-F. Lyu, {\it {W-boson mass, electroweak precision
  tests, and SMEFT}},  {\it Phys. Rev. D} {\bf 106} (2022), no.~7 073010,
  [\href{http://arxiv.org/abs/2204.04805}{{\tt arXiv:2204.04805}}].

\bibitem{Liu:2022jdq}
X.~Liu, S.-Y. Guo, B.~Zhu, and Y.~Li, {\it {Correlating Gravitational Waves
  with $W$-boson Mass, FIMP Dark Matter, and Majorana Seesaw Mechanism}},  {\it
  Sci. Bull.} {\bf 67} (2022) 1437--1442,
  [\href{http://arxiv.org/abs/2204.04834}{{\tt arXiv:2204.04834}}].

\bibitem{Lee:2022nqz}
H.~M. Lee and K.~Yamashita, {\it {A model of vector-like leptons for the muon
  $g-2$ and the W boson mass}},  {\it Eur. Phys. J. C} {\bf 82} (2022), no.~8
  661, [\href{http://arxiv.org/abs/2204.05024}{{\tt arXiv:2204.05024}}].

\bibitem{Cheng:2022jyi}
Y.~Cheng, X.-G. He, Z.-L. Huang, and M.-W. Li, {\it {Type-II seesaw triplet
  scalar effects on neutrino trident scattering}},  {\it Phys. Lett. B} {\bf
  831} (2022) 137218, [\href{http://arxiv.org/abs/2204.05031}{{\tt
  arXiv:2204.05031}}].

\bibitem{Song:2022xts}
H.~Song, W.~Su, and M.~Zhang, {\it {Electroweak phase transition in 2HDM under
  Higgs, Z-pole, and W precision measurements}},  {\it JHEP} {\bf 10} (2022)
  048, [\href{http://arxiv.org/abs/2204.05085}{{\tt arXiv:2204.05085}}].

\bibitem{Bagnaschi:2022whn}
E.~Bagnaschi, J.~Ellis, M.~Madigan, K.~Mimasu, V.~Sanz, and T.~You, {\it {SMEFT
  analysis of $m_{W}$}},  {\it JHEP} {\bf 08} (2022) 308,
  [\href{http://arxiv.org/abs/2204.05260}{{\tt arXiv:2204.05260}}].

\bibitem{Paul:2022dds}
A.~Paul and M.~Valli, {\it {Violation of custodial symmetry from W-boson mass
  measurements}},  {\it Phys. Rev. D} {\bf 106} (2022), no.~1 013008,
  [\href{http://arxiv.org/abs/2204.05267}{{\tt arXiv:2204.05267}}].

\bibitem{Bahl:2022xzi}
H.~Bahl, J.~Braathen, and G.~Weiglein, {\it {New physics effects on the W-boson
  mass from a doublet extension of the SM Higgs sector}},  {\it Phys. Lett. B}
  {\bf 833} (2022) 137295, [\href{http://arxiv.org/abs/2204.05269}{{\tt
  arXiv:2204.05269}}].

\bibitem{Asadi:2022xiy}
P.~Asadi, C.~Cesarotti, K.~Fraser, S.~Homiller, and A.~Parikh, {\it {Oblique
  Lessons from the $W$ Mass Measurement at CDF II}},
  \href{http://arxiv.org/abs/2204.05283}{{\tt arXiv:2204.05283}}.

\bibitem{DiLuzio:2022xns}
L.~Di~Luzio, R.~Gr\"ober, and P.~Paradisi, {\it {Higgs physics confronts the
  $M_W$ anomaly}},  {\it Phys. Lett. B} {\bf 832} (2022) 137250,
  [\href{http://arxiv.org/abs/2204.05284}{{\tt arXiv:2204.05284}}].

\bibitem{Athron:2022isz}
P.~Athron, M.~Bach, D.~H.~J. Jacob, W.~Kotlarski, D.~St\"ockinger, and
  A.~Voigt, {\it {Precise calculation of the W boson pole mass beyond the
  standard model with FlexibleSUSY}},  {\it Phys. Rev. D} {\bf 106} (2022),
  no.~9 095023, [\href{http://arxiv.org/abs/2204.05285}{{\tt
  arXiv:2204.05285}}].

\bibitem{Gu:2022htv}
J.~Gu, Z.~Liu, T.~Ma, and J.~Shu, {\it {Speculations on the W-mass measurement
  at CDF*}},  {\it Chin. Phys. C} {\bf 46} (2022), no.~12 123107,
  [\href{http://arxiv.org/abs/2204.05296}{{\tt arXiv:2204.05296}}].

\bibitem{Heckman:2022the}
J.~J. Heckman, {\it {Extra W-boson mass from a D3-brane}},  {\it Phys. Lett. B}
  {\bf 833} (2022) 137387, [\href{http://arxiv.org/abs/2204.05302}{{\tt
  arXiv:2204.05302}}].

\bibitem{Babu:2022pdn}
K.~S. Babu, S.~Jana, and V.~P. K., {\it {Correlating W-Boson Mass Shift with
  Muon g-2 in the Two Higgs Doublet Model}},  {\it Phys. Rev. Lett.} {\bf 129}
  (2022), no.~12 121803, [\href{http://arxiv.org/abs/2204.05303}{{\tt
  arXiv:2204.05303}}].

\bibitem{Heo:2022dey}
Y.~Heo, D.-W. Jung, and J.~S. Lee, {\it {Impact of the CDF W-mass anomaly on
  two Higgs doublet model}},  {\it Phys. Lett. B} {\bf 833} (2022) 137274,
  [\href{http://arxiv.org/abs/2204.05728}{{\tt arXiv:2204.05728}}].

\bibitem{Du:2022brr}
X.~K. Du, Z.~Li, F.~Wang, and Y.~K. Zhang, {\it {Explaining The New CDFII
  W-Boson Mass In The Georgi-Machacek Extension Models}},
  \href{http://arxiv.org/abs/2204.05760}{{\tt arXiv:2204.05760}}.

\bibitem{Cheung:2022zsb}
K.~Cheung, W.-Y. Keung, and P.-Y. Tseng, {\it {Iso-doublet Vector Leptoquark
  solution to the Muon $g-2$, $R_{K, K^*}$, $R_{D,D^*}$, and $W$-mass
  Anomalies}},  {\it Phys. Rev. D} {\bf 106} (2022), no.~1 015029,
  [\href{http://arxiv.org/abs/2204.05942}{{\tt arXiv:2204.05942}}].

\bibitem{Crivellin:2022fdf}
A.~Crivellin, M.~Kirk, T.~Kitahara, and F.~Mescia, {\it {Large $t\to cZ$ as a
  sign of vectorlike quarks in light of the W mass}},  {\it Phys. Rev. D} {\bf
  106} (2022), no.~3 L031704, [\href{http://arxiv.org/abs/2204.05962}{{\tt
  arXiv:2204.05962}}].

\bibitem{Endo:2022kiw}
M.~Endo and S.~Mishima, {\it {New physics interpretation of W-boson mass
  anomaly}},  {\it Phys. Rev. D} {\bf 106} (2022), no.~11 115005,
  [\href{http://arxiv.org/abs/2204.05965}{{\tt arXiv:2204.05965}}].

\bibitem{Biekotter:2022abc}
T.~Biek\"otter, S.~Heinemeyer, and G.~Weiglein, {\it {Excesses in the low-mass
  Higgs-boson search and the W-boson mass measurement}},
  \href{http://arxiv.org/abs/2204.05975}{{\tt arXiv:2204.05975}}.

\bibitem{Balkin:2022glu}
R.~Balkin, E.~Madge, T.~Menzo, G.~Perez, Y.~Soreq, and J.~Zupan, {\it {On the
  implications of positive W mass shift}},  {\it JHEP} {\bf 05} (2022) 133,
  [\href{http://arxiv.org/abs/2204.05992}{{\tt arXiv:2204.05992}}].

\bibitem{HFLAV:2019otj}
{\bf HFLAV} Collaboration, Y.~S. Amhis et~al., {\it {Averages of b-hadron,
  c-hadron, and $\tau $-lepton properties as of 2018}},  {\it Eur. Phys. J. C}
  {\bf 81} (2021), no.~3 226, [\href{http://arxiv.org/abs/1909.12524}{{\tt
  arXiv:1909.12524}}].

\bibitem{HFLAV:2022pwe}
{\bf HFLAV} Collaboration, Y.~Amhis et~al., {\it {Averages of $b$-hadron,
  $c$-hadron, and $\tau$-lepton properties as of 2021}},
  \href{http://arxiv.org/abs/2206.07501}{{\tt arXiv:2206.07501}}.

\bibitem{HFLAV:2022fall}
{\bf HFLAV} Collaboration, ``{Average of $R(D)$ and $R(D^\ast)$ for End of
  2022}.'' at
  \url{https://hflav-eos.web.cern.ch/hflav-eos/semi/fall22/html/RDsDsstar/RDRDs.html}.

\bibitem{Bigi:2016mdz}
D.~Bigi and P.~Gambino, {\it {Revisiting $B\to D \ell \nu$}},  {\it Phys. Rev.}
  {\bf D94} (2016), no.~9 094008, [\href{http://arxiv.org/abs/1606.08030}{{\tt
  arXiv:1606.08030}}].

\bibitem{Bernlochner:2017jka}
F.~U. Bernlochner, Z.~Ligeti, M.~Papucci, and D.~J. Robinson, {\it {Combined
  analysis of semileptonic $B$ decays to $D$ and $D^*$: $R(D^{(*)})$,
  $|V_{cb}|$, and new physics}},  {\it Phys. Rev.} {\bf D95} (2017), no.~11
  115008, [\href{http://arxiv.org/abs/1703.05330}{{\tt arXiv:1703.05330}}].
  [erratum: Phys. Rev.D97,no.5,059902(2018)].

\bibitem{Bigi:2017jbd}
D.~Bigi, P.~Gambino, and S.~Schacht, {\it {$R(D^*)$, $|V_{cb}|$, and the Heavy
  Quark Symmetry relations between form factors}},  {\it JHEP} {\bf 11} (2017)
  061, [\href{http://arxiv.org/abs/1707.09509}{{\tt arXiv:1707.09509}}].

\bibitem{Jaiswal:2017rve}
S.~Jaiswal, S.~Nandi, and S.~K. Patra, {\it {Extraction of $|V_{cb}|$ from
  $B\to D^{(*)}\ell\nu_\ell$ and the Standard Model predictions of
  $R(D^{(*)})$}},  {\it JHEP} {\bf 12} (2017) 060,
  [\href{http://arxiv.org/abs/1707.09977}{{\tt arXiv:1707.09977}}].

\bibitem{Bordone:2019vic}
M.~Bordone, M.~Jung, and D.~van Dyk, {\it {Theory determination of $\bar{B}\to
  D^{(*)}\ell^-\bar\nu$ form factors at $\mathcal{O}(1/m_c^2)$}},  {\it Eur.
  Phys. J. C} {\bf 80} (2020), no.~2 74,
  [\href{http://arxiv.org/abs/1908.09398}{{\tt arXiv:1908.09398}}].

\bibitem{Gambino:2019sif}
P.~Gambino, M.~Jung, and S.~Schacht, {\it {The $V_{cb}$ puzzle: An update}},
  {\it Phys. Lett. B} {\bf 795} (2019) 386--390,
  [\href{http://arxiv.org/abs/1905.08209}{{\tt arXiv:1905.08209}}].

\bibitem{BaBar:2019vpl}
{\bf BaBar} Collaboration, J.~P. Lees et~al., {\it {Extraction of form Factors
  from a Four-Dimensional Angular Analysis of $\overline{B} \rightarrow D^\ast
  \ell^- \overline{\nu}_\ell$}},  {\it Phys. Rev. Lett.} {\bf 123} (2019),
  no.~9 091801, [\href{http://arxiv.org/abs/1903.10002}{{\tt
  arXiv:1903.10002}}].

\bibitem{Martinelli:2021onb}
G.~Martinelli, S.~Simula, and L.~Vittorio, {\it {$\vert V_{cb} \vert$ and
  $R(D)^{(*)}$) using lattice QCD and unitarity}},  {\it Phys. Rev. D} {\bf
  105} (2022), no.~3 034503, [\href{http://arxiv.org/abs/2105.08674}{{\tt
  arXiv:2105.08674}}].

\bibitem{FermilabLattice:2021cdg}
{\bf Fermilab Lattice, MILC} Collaboration, A.~Bazavov et~al., {\it
  {Semileptonic form factors for $B\rightarrow D^*\ell \nu $ at nonzero recoil
  from $2+1$-flavor lattice QCD: Fermilab Lattice~and~MILC~Collaborations}},
  {\it Eur. Phys. J. C} {\bf 82} (2022), no.~12 1141,
  [\href{http://arxiv.org/abs/2105.14019}{{\tt arXiv:2105.14019}}]. [Erratum:
  Eur.Phys.J.C 83, 21 (2023)].

\bibitem{Bhaskar:2022vgk}
A.~Bhaskar, A.~A. Madathil, T.~Mandal, and S.~Mitra, {\it {Combined explanation
  of W-mass, muon g-2, RK(*) and RD(*) anomalies in a singlet-triplet scalar
  leptoquark model}},  {\it Phys. Rev. D} {\bf 106} (2022), no.~11 115009,
  [\href{http://arxiv.org/abs/2204.09031}{{\tt arXiv:2204.09031}}].

\bibitem{Li:2022gwc}
X.-Q. Li, Z.-J. Xie, Y.-D. Yang, and X.-B. Yuan, {\it {Correlating the CDF
  W-boson mass shift with the $b\to s\ell^+\ell^-$ anomalies}},  {\it Phys.
  Lett. B} {\bf 838} (2023) 137651,
  [\href{http://arxiv.org/abs/2205.02205}{{\tt arXiv:2205.02205}}].

\bibitem{Chowdhury:2022dps}
T.~A. Chowdhury and S.~Saad, {\it {Leptoquark-vectorlike quark model for the
  CDF $m_W$, $(g-2)_\mu$, $R_{K^{(\ast)}}$ anomalies, and neutrino masses}},
  {\it Phys. Rev. D} {\bf 106} (2022), no.~5 055017,
  [\href{http://arxiv.org/abs/2205.03917}{{\tt arXiv:2205.03917}}].

\bibitem{Allanach:2022bik}
B.~Allanach and J.~Davighi, {\it {$M_W$ helps select $Z^\prime$ models for
  $b\rightarrow s \ell \ell $ anomalies}},  {\it Eur. Phys. J. C} {\bf 82}
  (2022), no.~8 745, [\href{http://arxiv.org/abs/2205.12252}{{\tt
  arXiv:2205.12252}}].

\bibitem{Zheng:2022ssr}
M.-D. Zheng, F.-Z. Chen, and H.-H. Zhang, {\it {Explaining anomalies of
  B-physics, muon $g-2$ and W mass in R-parity violating MSSM with seesaw
  mechanism}},  {\it Eur. Phys. J. C} {\bf 82} (2022), no.~10 895,
  [\href{http://arxiv.org/abs/2207.07636}{{\tt arXiv:2207.07636}}].

\bibitem{Aaij:2020nrf}
{\bf LHCb} Collaboration, R.~Aaij et~al., {\it {Measurement of $CP$-Averaged
  Observables in the $B^{0}\rightarrow K^{*0}\mu^{+}\mu^{-}$ Decay}},  {\it
  Phys. Rev. Lett.} {\bf 125} (2020), no.~1 011802,
  [\href{http://arxiv.org/abs/2003.04831}{{\tt arXiv:2003.04831}}].

\bibitem{LHCb:2021zwz}
{\bf LHCb} Collaboration, R.~Aaij et~al., {\it {Branching Fraction Measurements
  of the Rare $B^0_s\rightarrow\phi\mu^+\mu^-$ and $B^0_s\rightarrow
  f_2^\prime(1525)\mu^+\mu^-$- Decays}},  {\it Phys. Rev. Lett.} {\bf 127}
  (2021), no.~15 151801, [\href{http://arxiv.org/abs/2105.14007}{{\tt
  arXiv:2105.14007}}].

\bibitem{LHCb:2022zom}
{\bf LHCb} Collaboration, {\it {Measurement of lepton universality parameters
  in $B^+\to K^+\ell^+\ell^-$ and $B^0\to K^{*0}\ell^+\ell^-$ decays}},
  \href{http://arxiv.org/abs/2212.09153}{{\tt arXiv:2212.09153}}.

\bibitem{LHCb:2017avl}
{\bf LHCb} Collaboration, R.~Aaij et~al., {\it {Test of lepton universality
  with $B^{0} \rightarrow K^{*0}\ell^{+}\ell^{-}$ decays}},  {\it JHEP} {\bf
  08} (2017) 055, [\href{http://arxiv.org/abs/1705.05802}{{\tt
  arXiv:1705.05802}}].

\bibitem{LHCb:2019hip}
{\bf LHCb} Collaboration, R.~Aaij et~al., {\it {Search for lepton-universality
  violation in $B^+\to K^+\ell^+\ell^-$ decays}},  {\it Phys. Rev. Lett.} {\bf
  122} (2019), no.~19 191801, [\href{http://arxiv.org/abs/1903.09252}{{\tt
  arXiv:1903.09252}}].

\bibitem{LHCb:2021trn}
{\bf LHCb} Collaboration, R.~Aaij et~al., {\it {Test of lepton universality in
  beauty-quark decays}},  {\it Nature Phys.} {\bf 18} (2022), no.~3 277--282,
  [\href{http://arxiv.org/abs/2103.11769}{{\tt arXiv:2103.11769}}].

\bibitem{Altmannshofer:2017poe}
W.~Altmannshofer, P.~S. Bhupal~Dev, and A.~Soni, {\it {$R_{D^{(*)}}$ anomaly: A
  possible hint for natural supersymmetry with $R$-parity violation}},  {\it
  Phys. Rev. D} {\bf 96} (2017), no.~9 095010,
  [\href{http://arxiv.org/abs/1704.06659}{{\tt arXiv:1704.06659}}].

\bibitem{Deshpand:2016cpw}
N.~G. Deshpande and X.-G. He, {\it {Consequences of R-parity violating
  interactions for anomalies in $\bar B\to D^{(*)} \tau \bar \nu$ and $b\to s
  \mu^+\mu^-$}},  {\it Eur. Phys. J.} {\bf C77} (2017), no.~2 134,
  [\href{http://arxiv.org/abs/1608.04817}{{\tt arXiv:1608.04817}}].

\bibitem{Trifinopoulos:2018rna}
S.~Trifinopoulos, {\it {Revisiting R-parity violating interactions as an
  explanation of the B-physics anomalies}},  {\it Eur. Phys. J.} {\bf C78}
  (2018), no.~10 803, [\href{http://arxiv.org/abs/1807.01638}{{\tt
  arXiv:1807.01638}}].

\bibitem{Altmannshofer:2020axr}
W.~Altmannshofer, P.~B. Dev, A.~Soni, and Y.~Sui, {\it {Addressing
  R$_{D^{(*)}}$, R$_{K^{(*)}}$, muon $g-2$ and ANITA anomalies in a minimal
  $R$-parity violating supersymmetric framework}},  {\it Phys. Rev. D} {\bf
  102} (2020), no.~1 015031, [\href{http://arxiv.org/abs/2002.12910}{{\tt
  arXiv:2002.12910}}].

\bibitem{Peskin:1990zt}
M.~E. Peskin and T.~Takeuchi, {\it {A New constraint on a strongly interacting
  Higgs sector}},  {\it Phys. Rev. Lett.} {\bf 65} (1990) 964--967.

\bibitem{Peskin:1991sw}
M.~E. Peskin and T.~Takeuchi, {\it {Estimation of oblique electroweak
  corrections}},  {\it Phys. Rev. D} {\bf 46} (1992) 381--409.

\bibitem{Heinemeyer:2004gx}
S.~Heinemeyer, W.~Hollik, and G.~Weiglein, {\it {Electroweak precision
  observables in the minimal supersymmetric standard model}},  {\it Phys.
  Rept.} {\bf 425} (2006) 265--368,
  [\href{http://arxiv.org/abs/hep-ph/0412214}{{\tt hep-ph/0412214}}].

\bibitem{Domingo:2011uf}
F.~Domingo and T.~Lenz, {\it {W mass and Leptonic Z-decays in the NMSSM}},
  {\it JHEP} {\bf 07} (2011) 101, [\href{http://arxiv.org/abs/1101.4758}{{\tt
  arXiv:1101.4758}}].

\bibitem{Heinemeyer:2013dia}
S.~Heinemeyer, W.~Hollik, G.~Weiglein, and L.~Zeune, {\it {Implications of LHC
  search results on the W boson mass prediction in the MSSM}},  {\it JHEP} {\bf
  12} (2013) 084, [\href{http://arxiv.org/abs/1311.1663}{{\tt
  arXiv:1311.1663}}].

\bibitem{Sirlin:1980nh}
A.~Sirlin, {\it {Radiative Corrections in the $SU(2)_L \times U(1)$ Theory: A
  Simple Renormalization Framework}},  {\it Phys. Rev. D} {\bf 22} (1980)
  971--981.

\bibitem{Marciano:1980pb}
W.~J. Marciano and A.~Sirlin, {\it {Radiative Corrections to Neutrino Induced
  Neutral Current Phenomena in the $SU(2)_L \times U(1)$ Theory}},  {\it Phys.
  Rev. D} {\bf 22} (1980) 2695. [Erratum: Phys.Rev.D 31, 213 (1985)].

\bibitem{Arnan:2019olv}
P.~Arnan, D.~Becirevic, F.~Mescia, and O.~Sumensari, {\it {Probing low energy
  scalar leptoquarks by the leptonic $W$ and $Z$ couplings}},  {\it JHEP} {\bf
  02} (2019) 109, [\href{http://arxiv.org/abs/1901.06315}{{\tt
  arXiv:1901.06315}}].

\bibitem{ParticleDataGroup:2020ssz}
{\bf Particle Data Group} Collaboration, P.~A. Zyla et~al., {\it {Review of
  Particle Physics}},  {\it PTEP} {\bf 2020} (2020), no.~8 083C01.

\bibitem{Earl:2018snx}
K.~Earl and T.~Grégoire, {\it {Contributions to $b \rightarrow s \ell \ell$
  Anomalies from $R$-Parity Violating Interactions}},  {\it JHEP} {\bf 08}
  (2018) 201, [\href{http://arxiv.org/abs/1806.01343}{{\tt arXiv:1806.01343}}].

\bibitem{ALEPH:2005ab}
{\bf ALEPH, DELPHI, L3, OPAL, SLD, LEP Electroweak Working Group, SLD
  Electroweak Group, SLD Heavy Flavour Group} Collaboration, S.~Schael et~al.,
  {\it {Precision electroweak measurements on the $Z$ resonance}},  {\it Phys.
  Rept.} {\bf 427} (2006) 257--454,
  [\href{http://arxiv.org/abs/hep-ex/0509008}{{\tt hep-ex/0509008}}].

\bibitem{Bryman:2021teu}
D.~Bryman, V.~Cirigliano, A.~Crivellin, and G.~Inguglia, {\it {Testing Lepton
  Flavor Universality with Pion, Kaon, Tau, and Beta Decays}},  {\it Ann. Rev.
  Nucl. Part. Sci.} {\bf 72} (2022) 69--91,
  [\href{http://arxiv.org/abs/2111.05338}{{\tt arXiv:2111.05338}}].

\bibitem{ATLAS:2021jyv}
{\bf ATLAS} Collaboration, G.~Aad et~al., {\it {Search for new phenomena in
  $pp$ collisions in final states with tau leptons, b-jets, and missing
  transverse momentum with the ATLAS detector}},  {\it Phys. Rev. D} {\bf 104}
  (2021), no.~11 112005, [\href{http://arxiv.org/abs/2108.07665}{{\tt
  arXiv:2108.07665}}].

\bibitem{ATLAS:2015oan}
{\bf ATLAS} Collaboration, G.~Aad et~al., {\it {Search for massive, long-lived
  particles using multitrack displaced vertices or displaced lepton pairs in pp
  collisions at $\sqrt{s}$ = 8 TeV with the ATLAS detector}},  {\it Phys. Rev.
  D} {\bf 92} (2015), no.~7 072004,
  [\href{http://arxiv.org/abs/1504.05162}{{\tt arXiv:1504.05162}}].

\bibitem{CMS:2016zgb}
{\bf CMS} Collaboration, V.~Khachatryan et~al., {\it {Searches for
  $R$-parity-violating supersymmetry in $pp $collisions at $\sqrt{s} =$ 8 TeV
  in final states with 0-4 leptons}},  {\it Phys. Rev. D} {\bf 94} (2016),
  no.~11 112009, [\href{http://arxiv.org/abs/1606.08076}{{\tt
  arXiv:1606.08076}}].

\bibitem{Bagnaschi:2022qhb}
E.~Bagnaschi, M.~Chakraborti, S.~Heinemeyer, I.~Saha, and G.~Weiglein, {\it
  {Interdependence of the new \textquotedblleft{}MUON G-2\textquotedblright{}
  result and the W-boson mass}},  {\it Eur. Phys. J. C} {\bf 82} (2022), no.~5
  474, [\href{http://arxiv.org/abs/2203.15710}{{\tt arXiv:2203.15710}}].

\bibitem{Heinemeyer:1998yj}
S.~Heinemeyer, W.~Hollik, and G.~Weiglein, {\it {FeynHiggs: A Program for the
  calculation of the masses of the neutral CP even Higgs bosons in the MSSM}},
  {\it Comput. Phys. Commun.} {\bf 124} (2000) 76--89,
  [\href{http://arxiv.org/abs/hep-ph/9812320}{{\tt hep-ph/9812320}}].

\bibitem{Heinemeyer:1998np}
S.~Heinemeyer, W.~Hollik, and G.~Weiglein, {\it {The Masses of the neutral CP -
  even Higgs bosons in the MSSM: Accurate analysis at the two loop level}},
  {\it Eur. Phys. J. C} {\bf 9} (1999) 343--366,
  [\href{http://arxiv.org/abs/hep-ph/9812472}{{\tt hep-ph/9812472}}].

\bibitem{Degrassi:2002fi}
G.~Degrassi, S.~Heinemeyer, W.~Hollik, P.~Slavich, and G.~Weiglein, {\it
  {Towards high precision predictions for the MSSM Higgs sector}},  {\it Eur.
  Phys. J. C} {\bf 28} (2003) 133--143,
  [\href{http://arxiv.org/abs/hep-ph/0212020}{{\tt hep-ph/0212020}}].

\bibitem{Frank:2006yh}
M.~Frank, T.~Hahn, S.~Heinemeyer, W.~Hollik, H.~Rzehak, and G.~Weiglein, {\it
  {The Higgs Boson Masses and Mixings of the Complex MSSM in the
  Feynman-Diagrammatic Approach}},  {\it JHEP} {\bf 02} (2007) 047,
  [\href{http://arxiv.org/abs/hep-ph/0611326}{{\tt hep-ph/0611326}}].

\bibitem{Hahn:2013ria}
T.~Hahn, S.~Heinemeyer, W.~Hollik, H.~Rzehak, and G.~Weiglein, {\it
  {High-Precision Predictions for the Light CP -Even Higgs Boson Mass of the
  Minimal Supersymmetric Standard Model}},  {\it Phys. Rev. Lett.} {\bf 112}
  (2014), no.~14 141801, [\href{http://arxiv.org/abs/1312.4937}{{\tt
  arXiv:1312.4937}}].

\bibitem{Bahl:2016brp}
H.~Bahl and W.~Hollik, {\it {Precise prediction for the light MSSM Higgs boson
  mass combining effective field theory and fixed-order calculations}},  {\it
  Eur. Phys. J. C} {\bf 76} (2016), no.~9 499,
  [\href{http://arxiv.org/abs/1608.01880}{{\tt arXiv:1608.01880}}].

\bibitem{Bahl:2017aev}
H.~Bahl, S.~Heinemeyer, W.~Hollik, and G.~Weiglein, {\it {Reconciling EFT and
  hybrid calculations of the light MSSM Higgs-boson mass}},  {\it Eur. Phys. J.
  C} {\bf 78} (2018), no.~1 57, [\href{http://arxiv.org/abs/1706.00346}{{\tt
  arXiv:1706.00346}}].

\bibitem{Bahl:2018qog}
H.~Bahl, T.~Hahn, S.~Heinemeyer, W.~Hollik, S.~Pa\ss{}ehr, H.~Rzehak, and
  G.~Weiglein, {\it {Precision calculations in the MSSM Higgs-boson sector with
  FeynHiggs 2.14}},  {\it Comput. Phys. Commun.} {\bf 249} (2020) 107099,
  [\href{http://arxiv.org/abs/1811.09073}{{\tt arXiv:1811.09073}}].

\end{thebibliography}\endgroup

\end{document}